\documentclass[intlimits,twoside,a4paper]{article}

\usepackage[cp1251]{inputenc}

\usepackage[eqsecnum]{cmpj3}

\issue{2021}{24}{4}{43601}
\doinumber{10.5488/CMP.24.43601}

\title[Phase separation dynamics in aqueous solutions of thermoresponsive polymers]
{Phase separation dynamics in aqueous solutions of thermoresponsive polymers}

\author[V. I. Kovalchuk]
{V.~I.~Kovalchuk}
\address{Taras Shevchenko National University of Kyiv, 64/13, Volodymyrska Str., 01601 Kyiv, Ukraine}

\Keywords{thermoresponsive polymer, spinodal decomposition, Cahn-Hilliard equation}

\date{Received April 14, 2021, in final form September 7, 2021}

\begin{document}
\maketitle
\begin{abstract}
Phase transition kinetics of aqueous hydroxypropyl cellulose solution was studied by using turbidimetric monitoring
and mathematical modelling techniques. Based on the nonlinear Cahn-Hilliard equation with a mobility depending
on the component concentration, the phase separation has been modeled on a simple one-dimensional Flory lattice.
For value set of the interfacial energy parameter, data were obtained on the changing of the average values
of the cluster sizes, their mass and concentration. The simulation results allow us to distinguish three stages of
the spinodal decomposition: early, intermediate and final. It was found that for the intermediate stage,
the kinetics of the cluster mass growth is described by a dependence that is characteristic of the usual diffusion
mass transfer; the change in the average cluster size can be represented by a scaling function with an exponent close
to 1/3, typical of the systems with a conserved scalar order parameter. It is shown that the concentration of clusters at
the final stage is determined by the temperature dependence of the interfacial energy.
\printkeywords
\end{abstract}

\section{\label{sec:introduction}Introduction}

Phase separation of thermoresponsive polymers --- cellulose ethers --- is an intensive area of researches based
on promising technologies in various fields, including the food industry, personal hygiene products, medicine,
pharmacology, and environmentally friendly materials~\cite{Anastas1998,Kamide2005}. The kinetics of phase
separation in solutions of cellulose derivatives is a rather complex process, the understanding of which is necessary
to create materials with specified physicochemical properties.

Aqueous solutions of thermoresponsive polymers undergo a sol-gel transition upon heating, returning to their
original state upon cooling~\cite{Garate2017,Lodge2018}. Gelation of such a system is associated with an increase
of the solution turbidity resulting from phase separation. To date, the gelation mechanism is still not well understood,
although many different hypotheses have been proposed (\cite{Fairclough2012} and references therein). Structural
and rheological properties of polymer gels based on cellulose derivatives have been intensively studied since 1935
and all the experimental data accumulated to date indicate that the mechanism of phase separation in such systems
is spinodal decomposition~\cite{Fairclough2012,Sarkar1979,Takeshita2010,Villetti2011}. Spinodal decomposition is the
initial stage of phase transformation when the system was preliminarily brought to a thermodynamically unstable state.
This instability at a given temperature corresponds to the section of the free energy curve where its second derivative
with respect to concentration is negative. In the region of spinodal decomposition, the solution domains with an
increased or decreased concentration relative to its average value --- clusters --- arising due to thermal fluctuations
become stable and begin to grow. This process is maintained by ascending diffusion~\cite{Skripov1979}, when the mutual
attraction of the same type of particles leads at the next time moments to an even greater increase of their concentration
in the cluster and to a further depletion of the adjacent solution zone. In the case of spinodal decomposition, the separation
of a substance into various phases occurs uniformly throughout the entire solution volume; therefore, the sol-gel
transition in such systems is also called the volume phase transition~\cite{Weibenborn2019,Xia2003}.

This paper is devoted to the study of the phase transition kinetics in aqueous solutions of thermoresponsive
polymers --- cellulose derivatives.

\section{Experiment}

As the object of our study we selected a 2~wt\% aqueous solution of hydroxypropyl cellulose manufactured by Alfa Aesar~\cite{HPCalfa}. For this hydroxypropyl cellulose, its molecular weight was $10^{5}$, and the substitution degree
was 75.7\%.

\begin{figure}[!t] 
\centering
\includegraphics [scale=0.42] {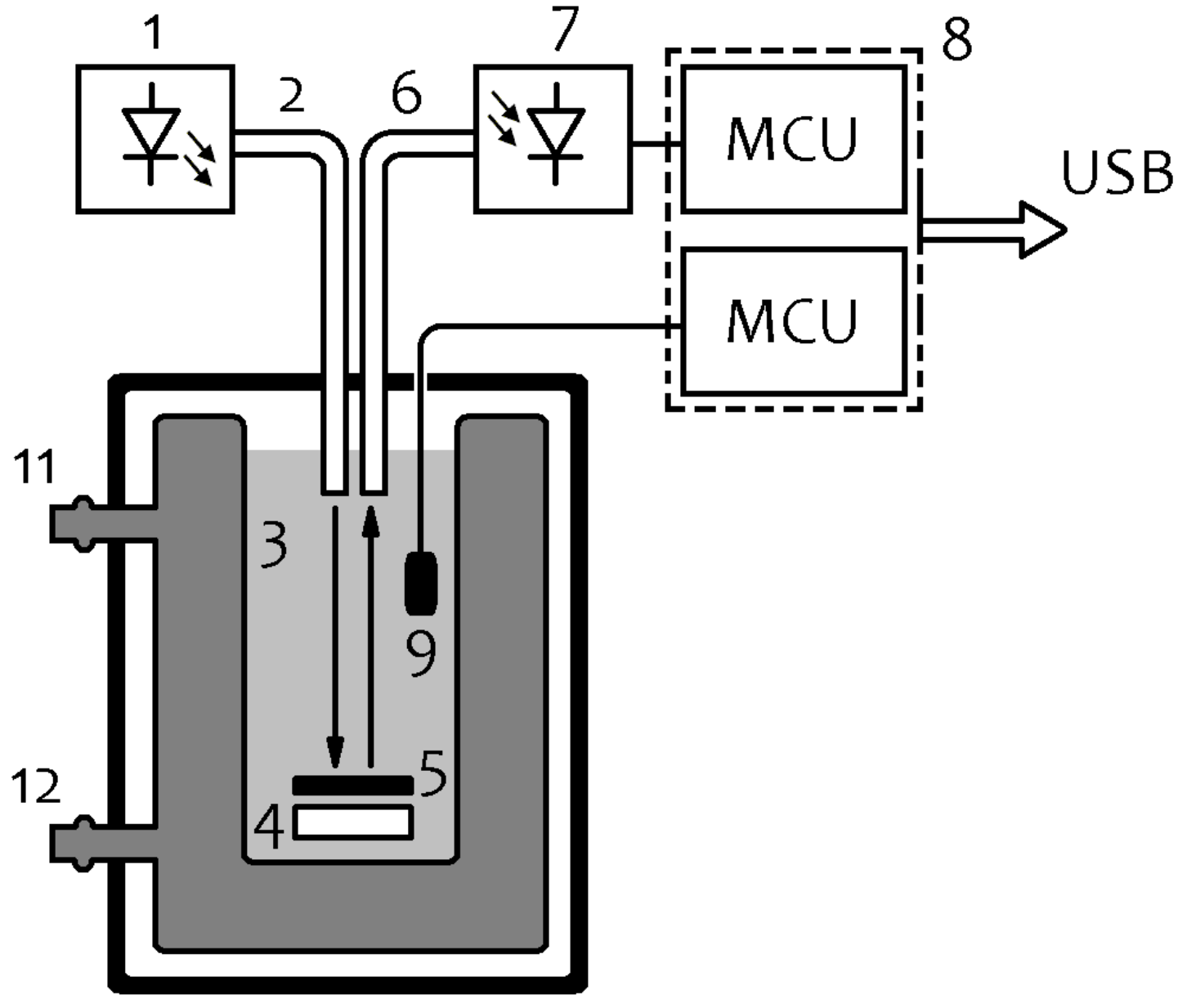}
\caption{Installation scheme for turbidity measuring (see explanations in the text).}
\label{fig1}
\end{figure}

Figure~\ref{fig1} illustrates the scheme of the experimental setup, which is an improved version of the device used
in~\cite{Alekseev2019} for turbidity measuring.

As a light source~1, the GNL-5013PGC LED was used. It was powered by a micropower current stabilizer made on
the LP2951 chip. Light beam with a wavelength of 525~nm was fed through the optical-fiber cable~2 into the
thermostated chamber~3 filled with a polymer solution. The beam reflected from the mirror~4 was returned via
the fiber cable~6 to the digital optical sensor~7 (TLS237) connected to the microcontroller~8 (AVR ATmega328P).
As a result, the total intensity was measured: \mbox{$J_{\Sigma}=J_{T}+J_{R}$}, where \mbox{$J_{T}$} is the
intensity of the weakened light flux due to the passage through the sample, \mbox{$J_{R}$} is the intensity
of the light flux reflected (backscattered) by the sample. To measure \mbox{$J_{R}$}, mirror~4 was shielded
by a shutter~5 made of light-absorbing material (black anodized aluminum). Thus, in order to determine
\mbox{$J_{T}$} and \mbox{$J_{R}$} for a specific sample, it was necessary to perform two measurements:
with and without screen~5.

The solution temperature was measured by the digital temperature sensor~9 (LMT01LPG) connected to the second
identical microcontroller of unit~8. The values of luminosity and temperature were synchronously read by the
unit~8 and transmitted via the RS-232 protocol into a USB ports of a PC. Data capture, their subsequent processing,
and visualization were implemented by a program written in Delphi. The luminosity measurement error did not exceed
0.1~lux, and the temperature measurement accuracy was \mbox{$\pm{0.1}^{\circ}$}C. Chamber~3 was enclosed in a
light-tight case~10 and was connected through outputs~11,12 to a Julabo ME-6 circulation thermostat. Structurally,
elements~4, 5, 9 together with the output (input) of fiber cables~2, 6 were designed as a probe  immersed
in the solution to be researched.

\begin{figure}[!t] 
\centering
\includegraphics [scale=0.45] {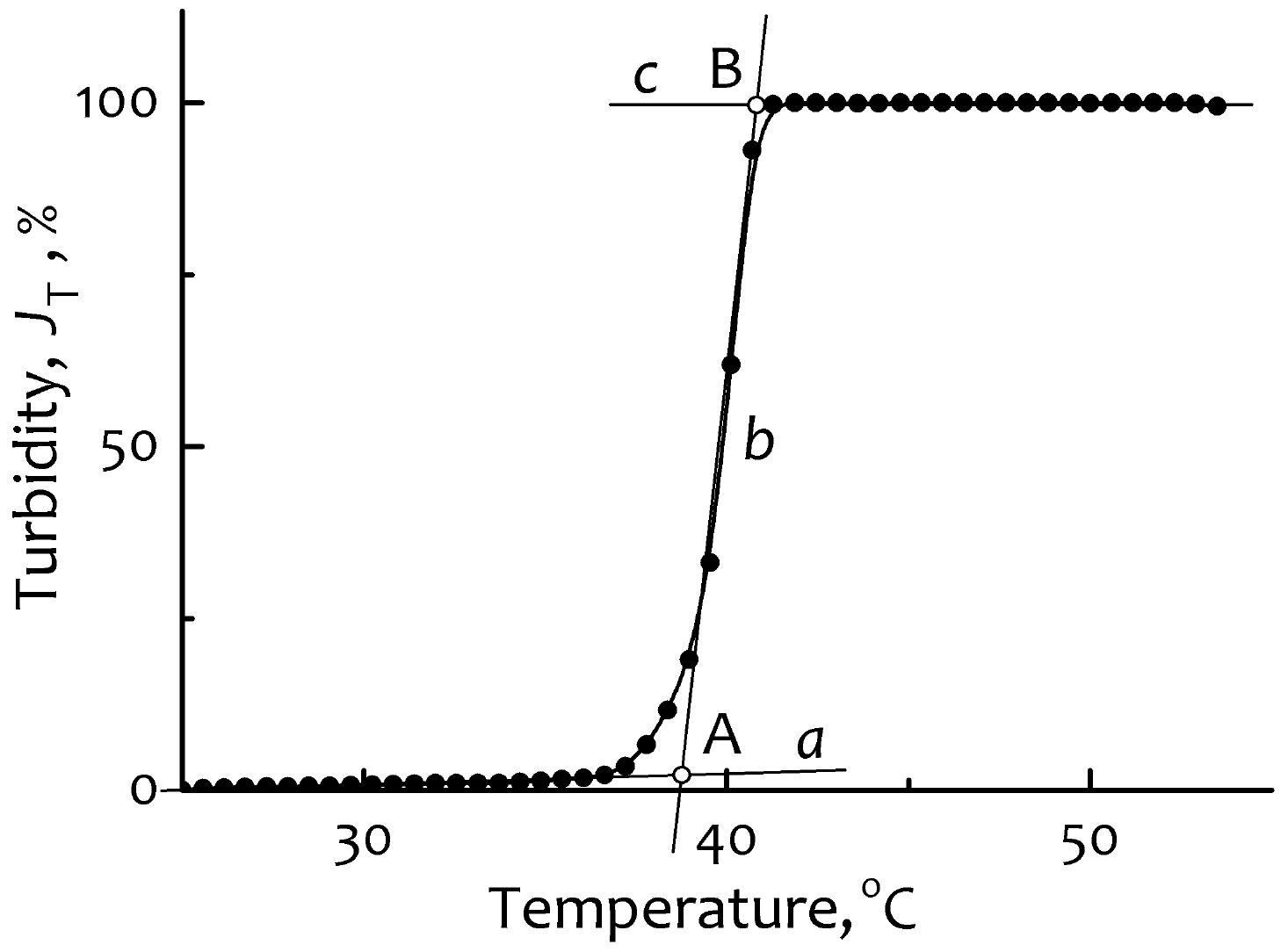}
\caption{Turbidity as a function of temperature for the 2~wt\% aqueous hydroxypropyl cellulose solution;
the temperature was ramped from 25 to \mbox{$60^{\circ}$}C at a rate of \mbox{$1.1^{\circ}$}C/min.
See other explanations in the text.}
\label{fig2}
\end{figure}

Figure~\ref{fig2} shows the measurement results for the investigated polymer solution. The curve in this figure was calibrated
with a turbidity level of 0\%, which corresponds to the illumination value of the photo sensor at a sample temperature
of \mbox{$25^{\circ}$}C.

From the analysis of the experimental curve behavior it follows that the observed turbidity can be conditionally divided
into three sections, each of which is approximated by linear dependencies (straight lines $a$, $b$, and $c$).
The intersection point of straight lines $a$ and $b$ corresponds to \mbox{$T_{A}=38.6^{\circ}$C} --- the so-called cloud
point~\cite{Fairclough2012}; the intersection point of straight lines $b$ and $c$ determines \mbox{$T_{B}=40.8^{\circ}$C}
--- the temperature of the phase transition completion. The phase transition temperature can be defined as
\mbox{$T_{P}=(T_{A}+T_{B})/2=39.7^{\circ}$}C. The obtained temperature is in agreement with the data of other authors:
\mbox{$T_{P}=39\pm1^{\circ}$}C~\cite{Desai2006}, \mbox{$T_{P}=40.3^{\circ}$}C~\cite{Khumana2014}.

\section{Formalism}

When describing the features of phase formation in the region of spinodal decomposition, one of the most successful
approaches is the Cahn-Hilliard model~\cite{Cahn1958,Cahn1961,Glotzer2002}, developed on the basis of the free energy
density functional method~\cite{Lee2014,L'vov2017}. The Cahn-Hilliard equation does not contain any microscopic details
of the described system, but includes such macroscopic characteristics as the diffusion coefficient, free and interfacial
energy. The free energy density functional method makes it possible to describe in a natural way the diffusion interaction
between clusters and their coagulation during the growth of a new phase without any approximations~\cite{L'vov2017}.

The Ginzburg-Landau type functional for the total energy of the polymer-solvent system has the form~\cite{Li2015}
\begin{equation}
U[\phi]=\int \rd{\bf{r}}
\Bigl\{
F(\phi)+\kappa(\phi)|\nabla\phi|^{2})
\Bigr\},
\quad
\phi=\phi({\bf{r}},t),
\label{eq1}
\end{equation}
where $\phi$ is the order parameter which has the meaning of the polymer concentration, $F(\phi)$ is the free energy
density. The second term in curly brackets under the integral sign in (\ref{eq1}) describes the contribution of spatial
correlation effects to free energy~\cite{Skripov1979} with the gradient coefficient~\cite{Gennes1980}
\begin{equation}
\kappa(\phi)=\frac{a^{2}}{36\,\phi(1-\phi)},
\label{eq2}
\end{equation}
where $a$ is the size of the polymer chain segments. For the Flory model, \mbox{$a=1$}~\cite{Flory1953}; therefore,
we omit this parameter in further notation. Let us introduce in (\ref{eq1}) the dimensionless parameter $\alpha$
\begin{equation}
U[\phi]=\int \rd{\bf{r}}
\Bigl\{
F(\phi)+\alpha^{2}\kappa(\phi)|\nabla\phi|^{2})
\Bigr\},
\label{eq3}
\end{equation}
the physical meaning of which will be clarified in the next section. For \mbox{$\alpha=1$}, the total energy has the
form (\ref{eq1}). In this work, we  consider the solutions of the Cahn-Hilliard equation with \mbox{$\alpha\ne1$}.

The Cahn-Hilliard equation describing the concentration evolution at a certain point in space ${\bf{r}}$
at a certain time moment $t$ has the form~\cite{Li2015}
\begin{equation}
\frac{\partial\phi}{\partial{t}}=\nabla\biggl\{\mathit{M}\nabla\frac{\delta{U[\phi]}}{\delta\phi}\biggr\}+\xi,
\label{eq4}
\end{equation}
where $\mathit{M}$ is the mobility, \mbox{$\xi=\xi({\bf{r}},t)$} is a stochastic function (thermal noise) that satisfies the
fluctuation-dissipation theorem~\cite{Landau2013}.

To solve equation (\ref{eq4}) with functional (\ref{eq3}), we use the Flory-Huggins free energy of mixing for
polymer-solvent systems~\cite{Flory1953,Huggins2013}
\begin{equation}
F(\phi)=N_{P}^{-1}\phi\ln\phi+(1-\phi)\ln(1-\phi)+\chi\phi(1-\phi),
\label{eq5}
\end{equation}
where $N_{P}$ is the polymerization degree, $\chi$ is the Flory-Huggins parameter describing pair interactions
between monomers. Then, the total energy variational derivative with respect to $\phi$ takes the form~\cite{Li2015}
\begin{equation}
\frac{\delta{U[\phi]}}{\delta\phi}=f(\phi)-\alpha^{2}\bigl[
\lambda(\phi)|\nabla\phi|^{2}+2\kappa(\phi)\Delta\phi\bigr],
\label{eq6}
\end{equation}
where
\begin{equation}
f(\phi)=N_{P}^{-1}(\ln\phi+1)-\ln(1-\phi)-\chi(2\phi-1)-1,
\label{eq7}
\end{equation}
\begin{equation}
\lambda(\phi)=\frac{2\phi-1}{36\,\phi^{2}(1-\phi)^{2}}.
\label{eq8}
\end{equation}
Let us believe that the mobility of polymer molecules depends on their concentration~\cite{Dolinnyi1994}
\begin{equation}
\mathit{M}=\mathit{M}_{0}\phi(1-\phi).
\label{eq9}
\end{equation}
Thus, the Cahn-Hilliard equation takes the form
%
\begin{equation}
\frac{\partial\phi}{\partial{t}}=\mathit{M}_{0}\nabla
 \left\lbrace 
\phi(1-\phi)\nabla
\left(
f(\phi)-\alpha^{2}
\bigl[
\lambda(\phi)|\nabla\phi|^{2}+2\kappa(\phi)\Delta\phi
\bigr]
\right)
\right\rbrace+\xi.
\label{eq10}
\end{equation}

This is a fourth-order nonlinear parabolic partial differential equation with a stochastic term. To enforce
the mass conservation law and descending of total energy with time, equation (\ref{eq10}) must be equipped
with the homogeneous Neumann boundary conditions~\cite{Lee2014}.

\section{Calculation results and discussion}

Equation (\ref{eq10}) describes the temporal evolution of a conservative (potential) field, which is a continuous
and sufficiently differentiable function. In~\cite{Chafee1974}, the solvability of the Cahn-Hilliard equation and
the existence of a set of stationary states, to which the initial distributions \mbox{$\phi({\bf{r}},0)$} are attracted
at \mbox{$t\rightarrow\infty$}, were proved.

Without loss of generality, let us investigate solutions of the one-dimensional equation (\ref{eq10}).
Introducing a spatio-temporal grid for \mbox{$x_{i}\in{[1,L]}$} \mbox{$(i=1,2,...,L)$} and
\mbox{$t^{n}\in{[0,\infty)}$} \mbox{$(n\in{\mathbb{N}})$} with periodic boundary conditions, we use for (\ref{eq10})
the semi-implicit difference scheme proposed in~\cite{Li2015}
\begin{equation}
\phi_{i}^{n+1}=\phi_{i}^{n}+{\mathit{M}}_{0}\tau\nabla_{h}
\Bigl\{
\phi_{i}^{n}(1-\phi_{i}^{n})\nabla_{h}
\left( 
f(\phi_{i}^{n})-\alpha^{2}
\bigl[
\lambda(\phi_{i}^{n})|\nabla_{h}\phi_{i}^{n}|^{2}+
2\kappa(\phi_{i}^{n})\Delta_{h}\phi_{i}^{n+1}
\bigr]
\right) 
\Bigr\}
+\epsilon\nabla_{h}\eta_{i}.
\label{eq11}
\end{equation}
Here, $\tau$ is the time step, \mbox{$h=1$} is the spatial step, $\nabla_{h}$ and $\Delta_{h}$ are the discrete versions
of corresponding differential operators~\cite{Samarskii2001}, $\epsilon$ is the intensity of thermal noise, \mbox{$\eta_{i}$}
is the sequence of standard normally distributed random variables, which is calculated once at the beginning of
computations.

According to (\ref{eq11}), the transition from the $n$-th time layer to the \mbox{$(n+1)$}-th time layer is carried out
in one step, but before that the value of \mbox{$\phi_{i}^{n+1}$} in the right-hand side of equation (\ref{eq11})
is found each time by the method of successive approximations from the initial \mbox{$\phi_{i}^{n}$}. The iterations
were performed until the condition
\begin{equation*}
\biggl|\frac{\phi_{i}^{n+1}(j+1)-\phi_{i}^{n+1}(j)}{\phi_{i}^{n+1}(j)}\biggr|\leqslant10^{-3}
\end{equation*}
was satisfied between two successive iterations $j$ and \mbox{$(j+1)$}, the maximum number of iterations did not exceed 10.

The Flory-Huggins interaction parameter is \mbox{$\chi=0.48$}~\cite{Chi} for the researched system, the initial phase was
homogeneous, and the corresponding polymer concentration for a 2~wt\% solution was \mbox{$\phi_{\text{in}}=0.02/\rho$},
where \mbox{$\rho=1.27\,\,\text{g/cm}^{3}$} is the density of hydroxypropyl cellulose~\cite{HPCdb}. The degree of polymerization
was chosen equal to \mbox{$N_{P}=300$}, the values of the other parameters were: \mbox{$\mathit{M}_{0}=0.3$}, \mbox{$\epsilon=10^{-5}$},
and \mbox{$\tau=10^{-5}$} (these optimal values are chosen exclusively for convenience reasons: maximum simulation time and
available computing resources, since parameters specified affect only the growth rate of clusters rather than the microscopic
features of their spatial structure). The phase structure growth was simulated on a lattice with \mbox{$L=128$}.

\begin{figure}[!t] 
\centering
\includegraphics [scale=0.95] {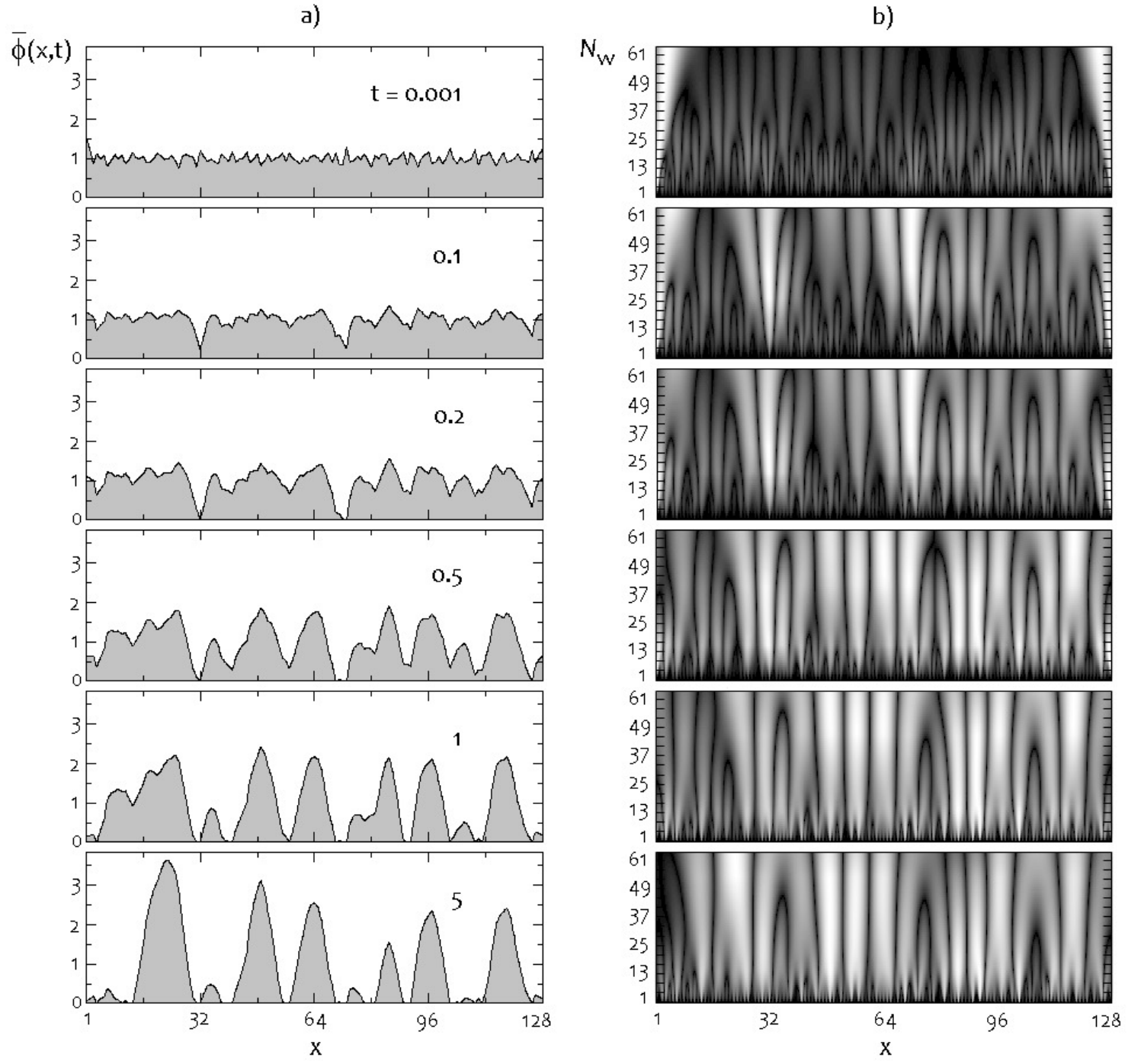}
\caption{Distribution of the reduced concentration \mbox{$\overline{\phi}(x,t)$} on a one-dimensional lattice during the
spinodal decomposition in the hydroxypropyl cellulose-water system (a); the result of the wavelet transforms of the
corresponding phase structures (b). The value of \mbox{$\alpha$} is 0.28.}
\label{fig3}
\end{figure}

In figure~\ref{fig3}a, as an example, the result of numerical simulation is demonstrated for \mbox{$\alpha=0.28$}, where it is seen how
the reduced concentration \mbox{$\overline{\phi}=\phi/\phi_{\text{in}}$} changes depending on the dimensionless distance $x$
and time $t$. It can be seen how an unstable regime arises on the basis of a random distribution \mbox{$(t=0.001-0.2)$},
which leads to the primary structure of a new phase \mbox{$(t=0.5)$}, and then to its further coarsening and final formation
\mbox{$(t=1-5)$}.

Figure~\ref{fig3}b shows the wavelet transforms of the corresponding spatio-temporal structures presented in figure~\ref{fig3}a;
MATLAB Wave Toolbox~\cite{Wavelet} was used to construct these pictures. The numbers of the expansion coefficients
using the Mexican Hat Wavelet, \mbox{$N_{W}$}, are shown on the vertical axis. In figure~\ref{fig3}b, it can be seen
that local features (non-smoothness) correspond to vertical lines going from the points where the singularities are
located. The picture of the wavelet coefficients reproduces the hierarchical structure of fluctuations in the value
of \mbox{$\overline{\phi}(x,t)$}: for the interval \mbox{$t=0.001-0.5$}, one can see how the distribution of
\mbox{$N_{W}(x,t)$} gradually loses its fractal features --- the characteristic branching inherent to local extrema
disappears over time.

\begin{figure}[!t] 
\centering
\includegraphics [scale=0.43] {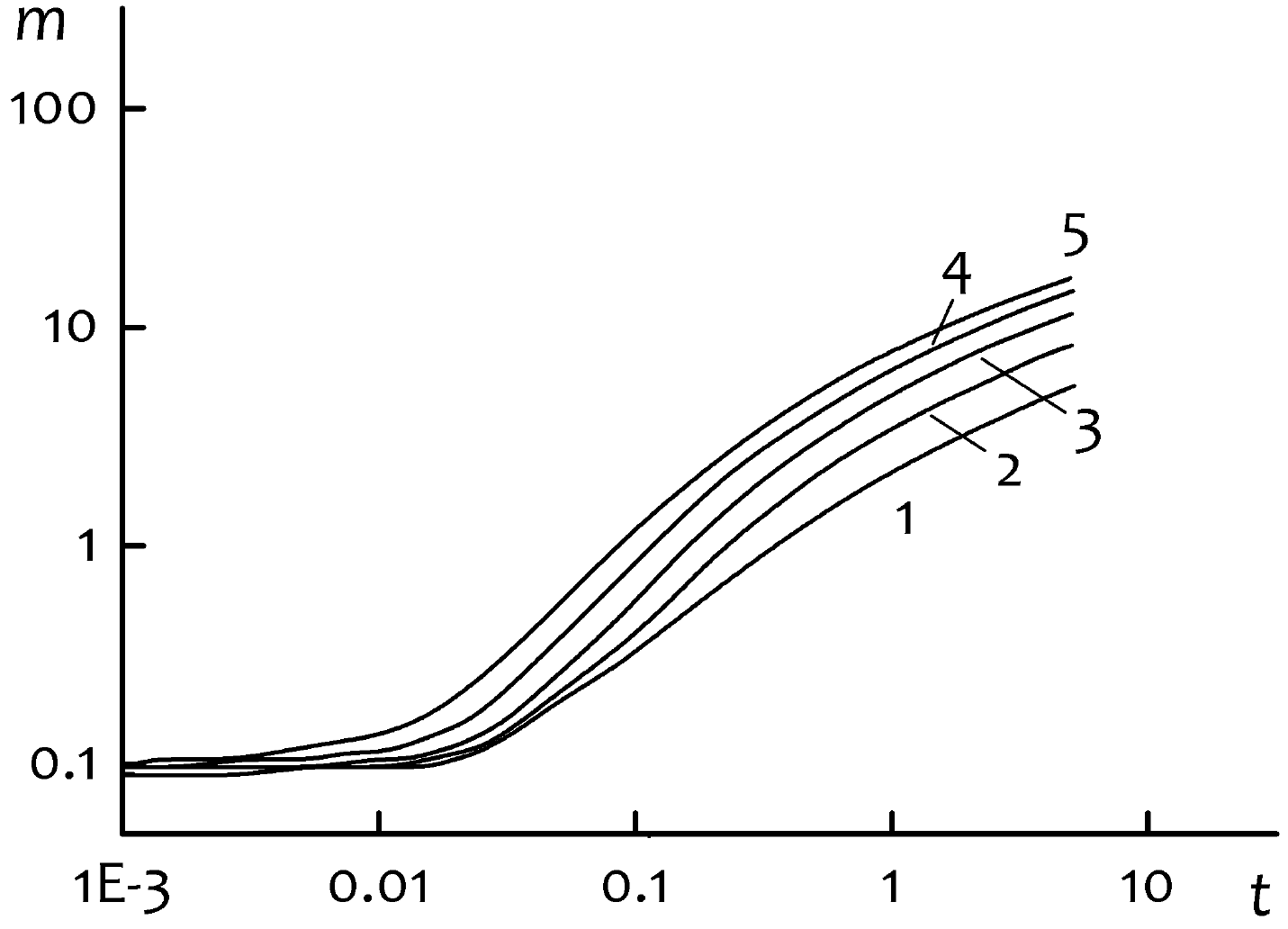}
\caption{Dependencies of the average cluster mass over time for the set of $\alpha$ values: 0.12~(1), 0.16~(2), 0.2~(3),
0.24~(4), 0.28~(5). Averaging was performed for 100 runs of the program simulating the phase structure.}
\label{fig4}
\end{figure}

Note that the self-organization process of the phase structure, modeled here by the Cahn-Hilliard equation, is provided
exclusively due to diffusion mass transfer, while other mechanisms are also possible, for example, mass transfer due to
interfacial tension~\cite{Siggia1979}. The mechanism of internal mass transfer governing the evolution of
\mbox{$\overline{\phi}(x,t)$} is confirmed by an analysis of the calculation results shown in figure~\ref{fig4}.
This figure shows the time dependencies of the average cluster mass \mbox{$m(t)=\langle{S/N_{c}}\rangle$}, where $S$ is
the area of the phase structure formed at the moment $t$ above level \mbox{$\overline{\phi}=1$}, and $N_{c}$ is the number of
clusters at the level \mbox{$\overline{\phi}=1$}. The kinetics of mass changing was analyzed using the power function
\mbox{$m(t)\sim{t^{\gamma}}$}: approximating the results in figure~\ref{fig4} by linear dependencies in the range \mbox{$t=0.5-5$},
and we get \mbox{$\gamma=0.545\pm{0.179}$}. This value of $\gamma$, close to 1/2, corresponds to the square root law for diffusion
mass transfer~\cite{Zhukhovitsky2001}.

\begin{figure}[!t] 
\centering
\includegraphics [scale=0.43] {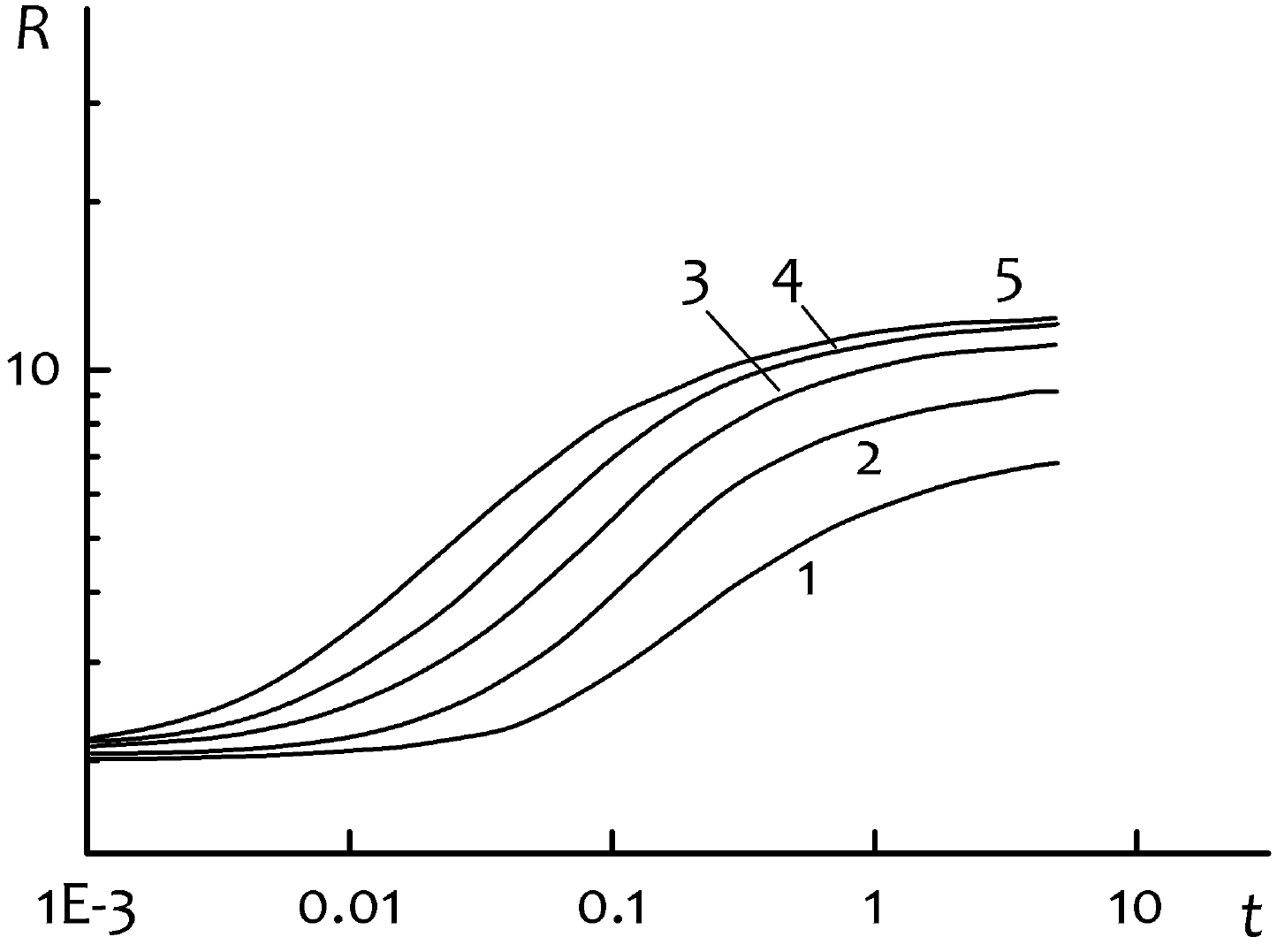}
\caption{The same as in figure~\ref{fig4}, but for the average cluster radius.}
\label{fig5}
\end{figure}
\begin{figure}[!t] 
\centering
\includegraphics [scale=0.43] {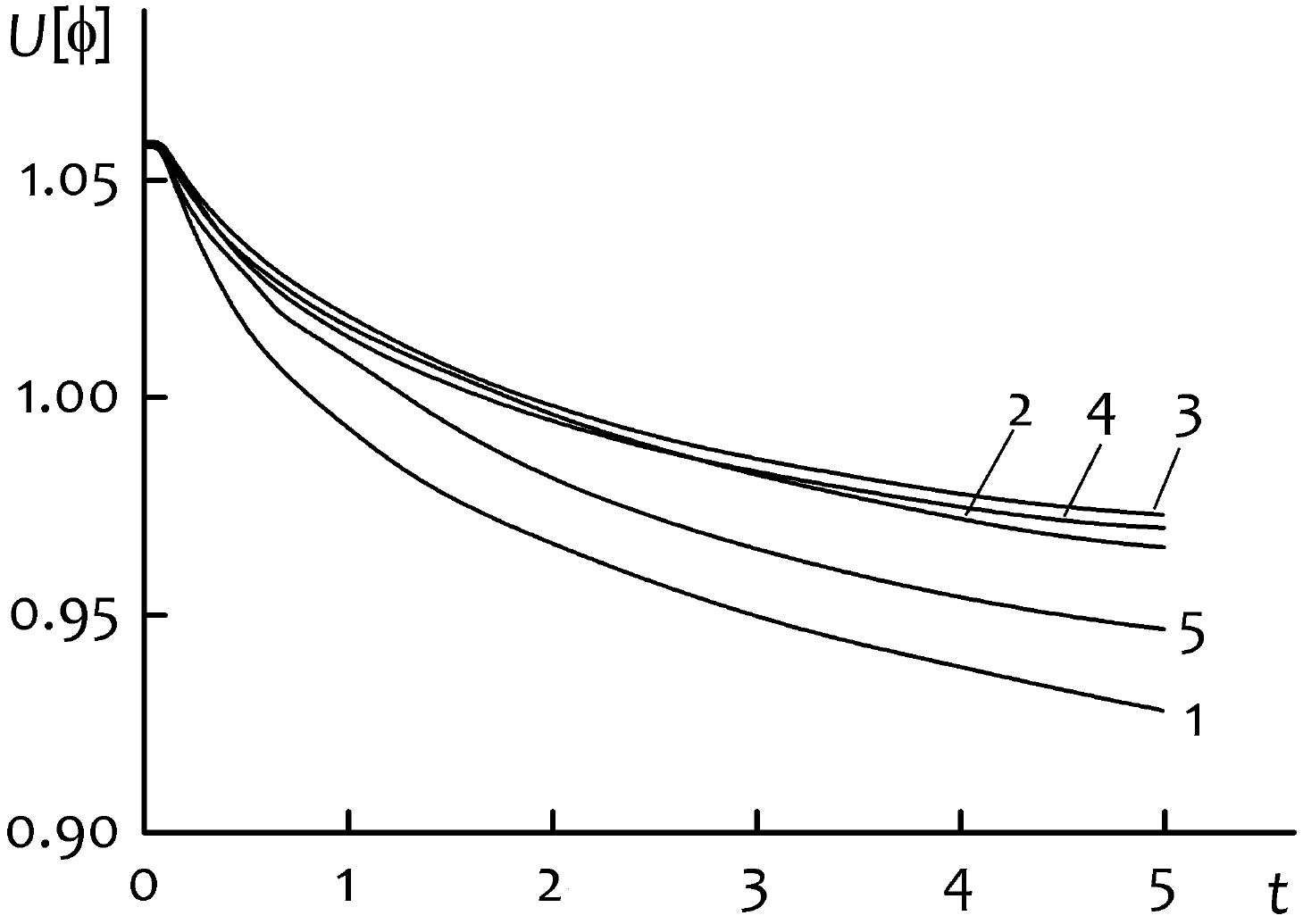}
\caption{The same as in figure~\ref{fig4}, but for the total energy.}
\label{fig6}
\end{figure}

Another characteristic of the asymptotic behavior of the system at large times is dynamic scaling which is determined by the
characteristic length $L(t)$ of separate ordered regions of the phase structure~\cite{Bray1994}. Choosing the average radius
of cluster $R(t)$ as characteristic length at the level \mbox{$\overline{\phi}=1$}, we find the scaling exponent for the growth
law \mbox{$R(t)\sim{t^{\delta}}$}. The analysis of the time dependencies $R(t)$ in their longest linear sections in figure~\ref{fig5}
gives the value \mbox{$\delta=0.307\pm{0.139}$}. This value is close to the exponent \mbox{$\delta=1/3$} which is typical
of the laws of growth in systems with a conserved scalar order parameter~\cite{Lifshitz1961,Bray1993}.

\begin{figure}[!t] 
\centering
\includegraphics [scale=0.43] {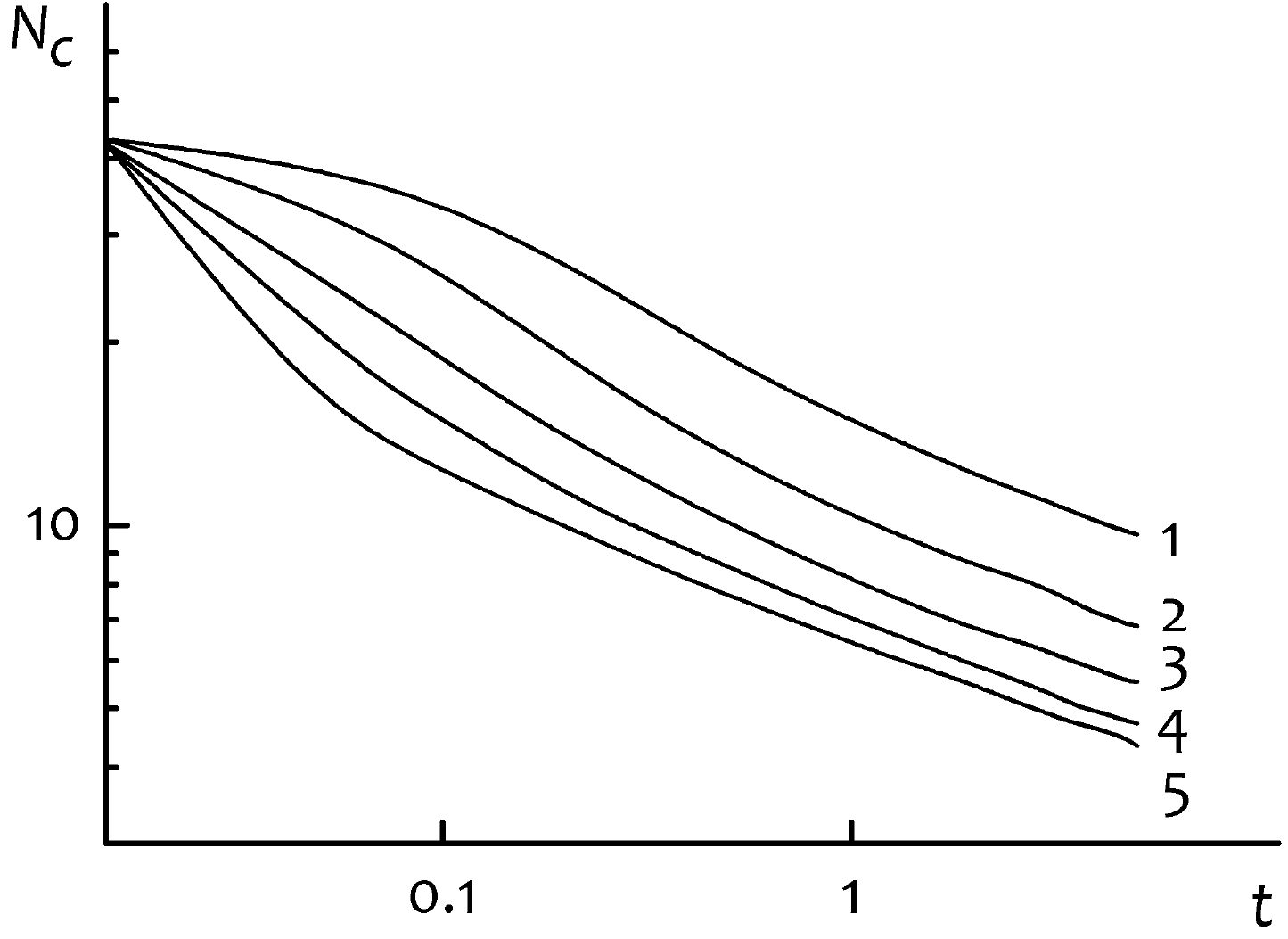}
\caption{The same as in figure~\ref{fig4}, but for the average number of clusters $N_{c}$.}
\label{fig7}
\end{figure}

The growth of the phase structure during spinodal decomposition occurs, on the whole, due to the coagulation of clusters
(as can be seen from figure~\ref{fig3}a), while the system relaxes into an energetically more favorable state with a lower energy
\mbox{$U[\phi]$}~(figure~\ref{fig6}).

The descent in the total energy over time, as follows from (\ref{eq3}), occurs not only due to the change in the free energy
$F(\phi)$, but also because of a decrease in the second term under the integral sign in (\ref{eq3}), which has a simple
physical meaning. Cahn and Hilliard~\cite{Cahn1958} showed that
\begin{equation}
\int \rd{\bf{r}}\,\kappa(\phi)|\nabla\phi|^{2}=\sigma/2\,,
\label{eq12}
\end{equation}
where $\sigma$ is the density of interfacial energy at equilibrium. During relaxation, the system tends to take a more
energetically favorable state, reducing the total surface area of the clusters due to their coagulation. As a result, the interfacial energy decreases. Consequently, the $\alpha$ parameter in (\ref{eq3}) determines the contribution
of the interfacial energy to the total energy; besides, $\alpha$ can also depend on temperature~\cite{Palmer1976,Lielmezs1986}
like the mobility $\mathit{M}_{0}$ or the Flory-Huggins parameter $\chi$. In the calculations, it was found that these three parameters
mainly determine the phase transition dynamics, but if $\mathit{M}_{0}$ and $\chi$ affect only the growth rate of new phase, then
$\alpha$ determines the number of clusters in the final stage of spinodal decomposition (figure~\ref{fig7}).

\begin{figure}[!t] 
\centering
\includegraphics [scale=0.43] {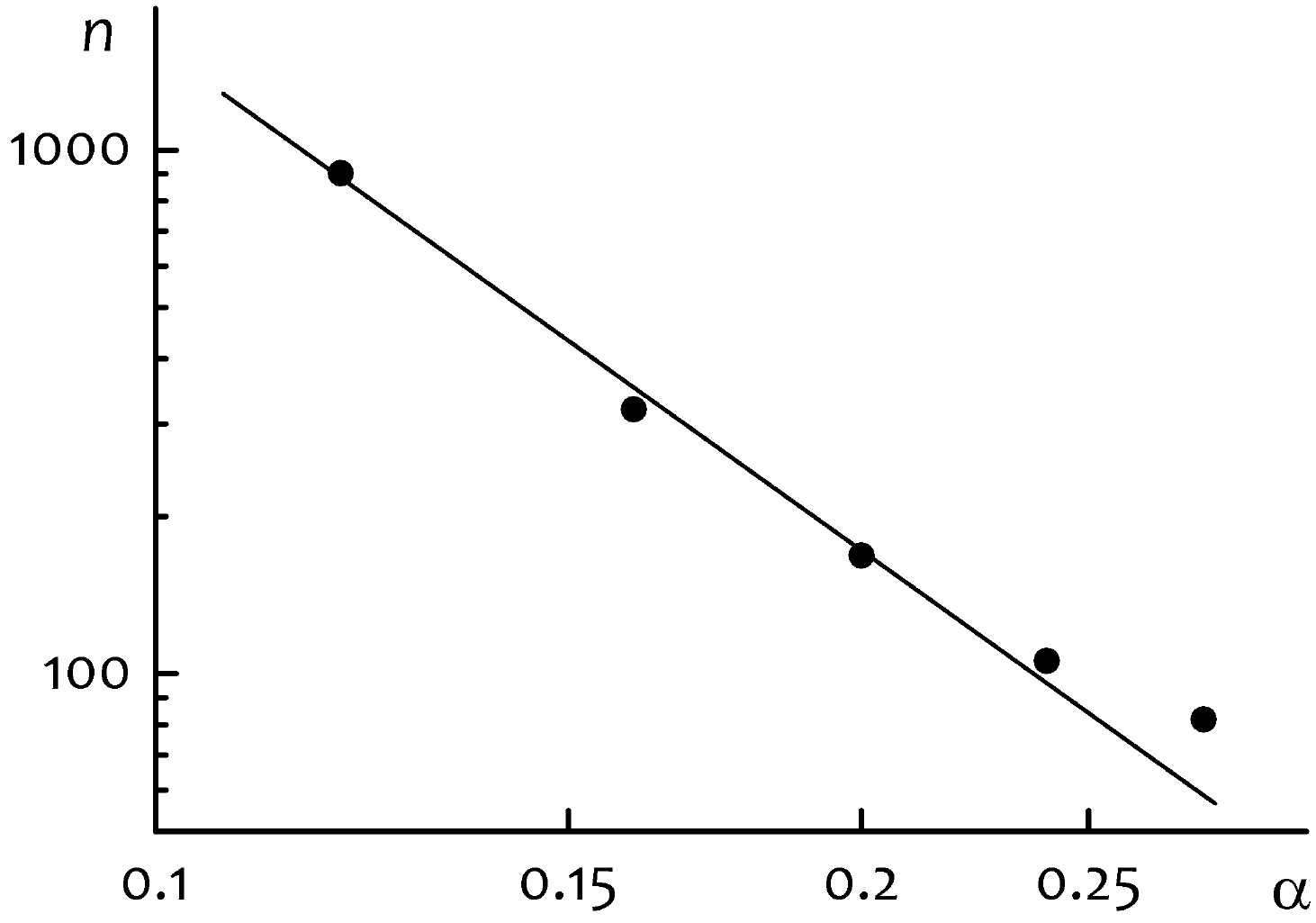}
\caption{Volume concentration of clusters $n$ as a function of parameter $\alpha$.
Simulation time is \mbox{$t=5$}.}
\label{fig8}
\end{figure}

In other words, if the density of interfacial energy depends on the temperature $T$, then the number of clusters 
also depends on $T$. It is possible to establish the form of the phenomenological dependence \mbox{$\alpha(T)$} based on
the following considerations.

The result shown in figure~\ref{fig7} was obtained for one-dimensional case. Assuming that the solution turbidity $J_{{T}}$
is proportional to the volume concentration of clusters \mbox{$n=N^{3}_{c}$}~\cite{Stratton1931}, we approximate the
dependence of \mbox{$n(\alpha)$} for the formed phase structure (at \mbox{$t=5$}) by the scaling function
\mbox{$n\sim\alpha^{\nu}$} with \mbox{$\nu=-(3.203\pm0.011)$} (figure~\ref{fig8}).

In the phase transition region, turbidity is proportional to temperature (section AB in figure~\ref{fig2}) or
\mbox{$J_{{T}}\sim\alpha^{\nu}\sim{T}$}, whence \mbox{$\alpha\sim{T^{-0.312}}$}, i.e.,
the contribution of the interfacial energy into the total energy decreases when temperature rises.

\section{Conclusion}

In the present study, the dynamic behavior of the phase separation in aqueous solution of hydroxypropyl cellulose was
analyzed by turbidimetry and mathematical modelling methods. Using the nonlinear Cahn-Hilliard equation with mobility
depending on the component concentration, the phase separation was simulated on a simple one-dimensional lattice.
For different values of the interfacial energy parameter, the time dependencies of the average values of cluster sizes,
their mass, and concentration per unit volume were obtained. The simulation results allow us to distinguish three stages
of the spinodal decomposition: early, intermediate, and final. For the intermediate stage, the kinetics of the change in
the average cluster mass is described by the dependence \mbox{$m(t)\sim{t^{\gamma}}$}, \mbox{$\gamma=0.545\pm{0.179}$},
which corresponds to the square root law typical of the usual diffusion mass transfer. The growth of the average size of the
phase structure elements can be described by the scaling function \mbox{$R(t)\sim{t^{\delta}}$} with the exponent
\mbox{$\delta=0.307\pm{0.139}$}; this value is close to 1/3 that is characteristic of the systems with a conserved scalar
order parameter. For the final stage, the concentration of clusters is determined by the contribution of the interfacial energy into
the total energy of the system; this contribution is described by the temperature dependence \mbox{$\alpha(T)\sim{T^{-0.312}}$}.

\section{Acknowledgements}

This work was supported by the Ministry of Education and Science of Ukraine by the grant ``Fractal dimension and properties
of liquid systems relevant for medicine and nuclear technologies'' (project No.~0120U102176).

All the simulations for this work were performed using the computing cluster of the Taras Shevchenko National University
of Kyiv~\cite{clusterKNU}.


\ukrainianpart

\title{Динаміка фазового розділення у водних розчинах термореактивних полімерів}
\author{В.~І.~Ковальчук}
\address{
 Київський національний університет імені Тараса Шевченка,
\\ вул. Володимирська, 64/13, Київ 01601, Україна}

\makeukrtitle
\begin{abstract}
\tolerance=3000%
Досліджена кінетика фазового переходу у водному розчині гідроксипропілцелюлози з використанням турбідиметричного
експерименту та методу математичного моделювання. На основі нелінійного рівняння Кана-Хілларда з рухливістю,
що залежить від концентрації компонента, виконана симуляція розділення фаз на простій одновимірній решітці Флорі.
Для набору значень параметра міжфазної енергії одержано дані про зміну у часі середніх розмірів кластерів, їх маси
та концентрації. Результати моделювання дозволяють виділити три стадії спінодального розпаду: ранню, проміжну та
фінальну. Встановлено, що для проміжної стадії кінетика збільшення маси кластера описується залежністю, характерною
для звичайної дифузійної масопередачі; зміна середнього розміру кластера може бути представлена скейлінговою функцією
з показником, близьким до 1/3, типовим для систем з консервативним скалярним параметром порядку. Показано, що
концентрація кластерів на фінальній стадії визначається температурною залежністю міжфазної енергії.
\keywords термореактивний полімер, спінодальний розпад, рівняння Кана-Хілларда
\end{abstract}


\begin{thebibliography}{99}

\bibitem{Anastas1998}
Anastas~P. T., Warner~J. C.,
Green Chemistry: Theory and Practice, Oxford University Press, New York, 1998.

\bibitem{Kamide2005}
Kamide~K.,
Cellulose and Cellulose Derivatives, Elsevier Science, Amsterdam, 2005.

\bibitem{Garate2017}
Garate~H., Li~K.-Wo, Bouyer~D., Guenoun~P.,
Soft Matter, 2017, \textbf{13}, 7161--7171, \doi{10.1039/c7sm01501a}.

\bibitem{Lodge2018}
Lodge~T. P., Maxwell~A.  L., Lott~J. R., Schmidt~P. W., McAllister~J. W., Morozova~S., Bates~F. S., Li~Y., Sammler~R. L.,
Biomacromolecules, 2018, \textbf{19}, No.~3, 816--824, \doi{10.1021/acs.biomac.7b01611}.

\bibitem{Fairclough2012}
Fairclough~J. P. A., Yu~H., Kelly~O., Ryan~A. J., Sammler~R. L., Radler~M.,
Langmuir, 2012, \textbf{28}, \\10551--10557, \doi{10.1021/la300971r}.

\bibitem{Sarkar1979}
Sarkar~N.,
J. Appl. Polym. Sci., 1979, \textbf{24}, 1073--1087, \doi{10.1002/app.1979.070240420}.

\bibitem{Takeshita2010}
Takeshita~H., Saito~K., Miya~M., Takenaka~K., Shiomi~T.,
J. Polym. Sci. B., 2010, \textbf{48}, 168--174, \\\doi{10.1002/polb.21885}.

\bibitem{Villetti2011}
Villetti~M. A., Soldi~V., Rochas~C., Borsali~R.,
Macromol. Chem. Phys., 2011, \textbf{212}, 1063--1071, \\\doi{10.1002/macp.201000697}.

\bibitem{Skripov1979}
Skripov~V. P., Skripov~A. V.,
Sov. Phys. Usp., 1979, \textbf{22}, 389--410, \doi{10.1070/PU1979v022n06ABEH005571}.

\bibitem{Weibenborn2019}
Wei{\ss}enborn~E., Braunschweig~B.,
Soft Matter, 2019, \textbf{15}, 2876--2883, \doi{10.1039/c9sm00093c}.

\bibitem{Xia2003}
Xia~X., Tang~S., Lu~X., Hu~Z.,
Macromolecules, 2003, \textbf{36}, 3695--3698, \doi{10.1021/ma0216728}.

\bibitem{HPCalfa}
Hydroxypropyl Cellulose,
URL~\url{https://www.alfa.com/en/catalog/043400/}.

\bibitem{Alekseev2019}
Alekseev~O. M., Zabashta~Yu. F., Kovalchuk~V. I., Lazarenko~M. M., Bulavin~L. A.,
Ukr. J. Phys., 2019, \textbf{64}, No.~3, 238--244, \doi{10.15407/ujpe64.3.238}.

\bibitem{Desai2006}
Desai~D., Rinaldi~F., Kothari~S., Paruchuri~S., Li~D., Lai~M., Fung~S., Both~D.,
Int. J. Pharm., 2006, \textbf{308}, 40--45, \doi{10.1016/j.ijpharm.2005.10.011}.

\bibitem{Khumana2014}
Khumana~P., Singh~W. B. K., Devi~S. D., Naorem~H.,
J. Macromol. Sci. A., 2014, \textbf{51}, 924--930, \\\doi{10.1080/10601325.2014.953377}.

\bibitem{Cahn1958}
Cahn~J. W., Hilliard~J. E.,
J. Chem. Phys., 1958, \textbf{28}, 258--267, \doi{10.1063/1.1744102}.

\bibitem{Cahn1961}
Cahn~J., Acta Metall., 1961, \textbf{9}, 795--801, \doi{10.1016/0001-6160(61)90182-1}.

\bibitem{Glotzer2002}
Glotzer~S. C., Paul~W.,
Annu. Rev. Mater. Res., 2002, \textbf{32}, 401--436, \\\doi{10.1146/annurev.matsci.32.010802.112213}.

\bibitem{Lee2014}
Lee~D., Huh~J.-Y., Jeong~D., Shin~J., Yun~A., Kim~J.,
Comput. Mater. Sci., 2014, \textbf{81}, 216--225, \\\doi{10.1016/j.commatsci.2013.08.027}.

\bibitem{L'vov2017}
L'vov~P. E., Svetukhin~V. V.,
Phys. Solid State, 2017, \textbf{59}, 355--361, \doi{10.1134/S1063783417020160}.

\bibitem{Li2015}
Li~X., Ji~G., Zhang~H.,
J. Comput. Phys., 2015, \textbf{283}, 81--97, \doi{10.1016/j.jcp.2014.11.032}.

\bibitem{Gennes1980}
De~Gennes~P. G.,
J. Chem. Phys., 1980, \textbf{72}, 4756--4763, \doi{10.1063/1.439809}.

\bibitem{Flory1953}
Flory~P. J.,
Principles of Polymer Chemistry, Cornell University Press, New York, 1953.

\bibitem{Landau2013}
Landau L. D., Lifshits E. M., Statistical Physics, Course of Theoretical Physics, Vol. 5, Elsevier, 3 edn., 2013.

\bibitem{Huggins2013}
Huggins~M. L.,
Physical Chemistry of High Polymers, Literary Licensing LLC, 2013.

\bibitem{Dolinnyi1994}
Dolinnyi~A. I.,
Polym. Sci. A., 1994, \textbf{36}, No.~5, 801--821 (in Russian).

\bibitem{Chafee1974}
Chafee~N., Infante~E. F.,
J. Appl. Anal., 1974, \textbf{4}, 17--37, \doi{10.1080/00036817408839081}.

\bibitem{Samarskii2001}
Samarskii~A. A.,
The Theory of Difference Schemes, CRC Press, New York, 2001.

\bibitem{Chi}
Polymer-Solvent Interaction Parameter at Infinite Dilution (Flory-Huggins),\\
URL~\url{http://polymerdatabase.com/polymer%20physics/Chi%20Table.html}.

\bibitem{HPCdb}
Hydroxypropyl Cellulose 9004-64-2,
URL~\url{https://www.guidechem.com/dictionary/en/9004-64-2.html}.

\bibitem{Wavelet}
Wavelet Toolbox,
URL~\url{https://uk.mathworks.com/help/wavelet/}.

\bibitem{Siggia1979}
Siggia~E. D.,
Phys. Rev. A, 1979, \textbf{20}, No.~2, 595--605, \doi{10.1103/PhysRevA.20.595}.

\bibitem{Zhukhovitsky2001}
Zhukhovitsky~A. A., Schwarzman~L. A.,
Physical Chemistry, Metallurgy, Moscow, 2001 (in Russian).

\bibitem{Bray1994}
Bray~A. J., Rutenberg~A. D.,
Phys. Rev. E, 1994, \textbf{49}, R27--R30, \doi{10.1103/physreve.49.r27}.

\bibitem{Lifshitz1961}
Lifshitz~I. M., Slyozov~V. V.,
J. Phys. Chem. Solids, 1961, \textbf{19}, 35--50, \doi{10.1016/0022-3697(61)90054-3}.

\bibitem{Bray1993}
Bray~A. J.,
Phys. Rev. E, 1993, \textbf{47}, 3191--3195, \doi{10.1103/PhysRevE.47.3191}.

\bibitem{Palmer1976}
Palmer~S. J.,
Phys. Educ., 1976, \textbf{11}, No.~2, 119--120, \doi{10.1088/0031-9120/11/2/009}.

\bibitem{Lielmezs1986}
Lielmezs~J., Herrick~T. A.,
Chem. Eng. J., 1986, \textbf{32}, 165--169, \doi{10.1016/0300-9467(86)80004-1}.

\bibitem{Stratton1931}
Stratton~J. A., Houghton~H. G.,
Phys. Rev. 1931, \textbf{38}, 159--165, \doi{10.1103/PhysRev.38.159}.

\bibitem{clusterKNU}
High-performance computing cluster of Information and Computer Center
of National Taras Shevchenko University of Kyiv,
URL~\url{http://cluster.univ.kiev.ua/eng/}.

\end{thebibliography}
\end{document}